\documentclass[twoside,fleqn]{article}
\usepackage{espcrc2,epsfig}

\newcommand{\AmS}{{\protect\the\textfont2
  A\kern-.1667em\lower.5ex\hbox{M}\kern-.125emS}}
\title{\rightline{\small  HUB--EP--96/34}
  \vspace*{-3mm}
  \rightline{\small July 26, 1996} 
    Physics of the Electroweak Phase Transition  at $M_H \leq 70$ GeV
    in a 3--dimensional $SU(2)$--Higgs Model 
}
\author{
    M.~G\"urtler$^1$, E.-M.~Ilgenfritz$^2$, J.~Kripfganz$^3$, H.~Perlt$^1$
    and A.~Schiller$^1$\thanks{Contribution presented by A. Schiller} \\
{\it $^1$ Institut f\"ur Theoretische Physik, Universit\"at Leipzig, Germany} \\
{\it $^2$ Institut f\"ur Physik, Humboldt-Universit\"at zu Berlin, Germany}\\
{\it $^3$ Institut f\"ur Theoretische Physik, Universit\"at Heidelberg, Germany}
}
\begin{document}
\begin{abstract}
 Physical parameters of the electroweak phase transition in a $3d$ effective
 lattice $SU(2)$--Higgs model are presented.
 The phase  transition temperatures, latent heats and continuum condensate 
 discontinuities are measured at Higgs masses of about 70 and 35 GeV.
 Masses and Higgs condensates are compared to perturbation theory in the broken
 phase. In the symmetric phase bound states and the static force are
 determined.
\end{abstract}

\maketitle
\section{Introduction}
\vspace*{-2mm}
In this contribution some of the results   given in \cite{guertler96} are
presented. 

Recent lattice studies of the electroweak phase transition
\cite{BunkEA}-\cite{physlett} had been triggered by the interest in
understanding BAU. The present quantitative understanding of possible 
mechanisms as well as the lower bounds for $M_H$ make BAU unlikely within 
the minimal standard model. Extensions, in particular supersymmetric ones, may
still be viable, however.

A second reason for lattice investigations was the wish to control the
behavior of perturbative calculations of the effective action. This
quantity is the appropriate tool of (non--lattice) thermal quantum
field theory for dealing with symmetry breaking. Infrared problems
prevent a perturbative evaluation of the free energy in the symmetric phase
to higher loops.

The  $SU(2)$ gauge--Higgs model has become a test-field to control the 
validity of perturbative predictions over a broad range of Higgs masses.  
Lattice simulations   make it possible to put both phases into coexistence
near the phase equilibrium. Thus one is able to measure directly physical
quantities quantifying the strength of the transition.

Furthermore, the $3d$ effective model  can be related to different $4d$ 
theories, {\it e.g.} with top.
\vspace*{-5mm}
\section{The strength of the phase transition}
\vspace*{-2mm} 
The action of the studied model is given by \cite{guertler96}
 \begin{eqnarray*}
S & = & \beta_G \sum_p \big(1 - {1 \over 2} Tr U_p \big) \\ 
  & - &   \beta_H \sum_l
       {1\over 2} Tr (\Phi_x^+ U_{x, \alpha} \Phi_{x + \alpha}) \\
 &+& \sum_x  \big( \rho_x^2 + \beta_R (\rho_x^2-1)^2 \big)
 \end{eqnarray*}
 with
\begin{eqnarray*}
\beta_G  =  \frac{4}{a g_3^2}, \ \  
\beta_R = \frac{\lambda_3}{g_3^2}  \ \frac{\beta_H^2}{\beta_G}, \\ 
\beta_H  = \frac{2 (1-2\beta_R)}{6+a^2 m_3^2}, \ \ \ 
   \frac{\lambda_3}{g_3^2}=\frac18 (\frac{M_H^*}{80\; \mbox{GeV}})^2 .
\end{eqnarray*}
In Table~1 the results for the phase transition parameters
are collected and tested on lattice spacing dependence 
 at $M_H^*=70$ GeV.
\begin{table}[!htb]
\begin{tabular*}{75mm } {@{\extracolsep{\fill}}|c|rrr| }
\hline
   ($\beta_G,M_H^*$)  & $\beta_{Hc}$& $T_c$ & $M_H$   \\
\hline   
(12,70)  &0.3435443(6)& 150.94(1)   &   64.77         \\
(16,70)  &0.3407942(6)& 151.27(2)   &   64.77         \\
(12,35)  &0.34140     & 76.2(1)     &   29.50         \\
\hline
\end{tabular*}
{ Table~1. \small \sl   $T_c$ and $4d$ 
Higgs masses (in $\mbox{GeV}$) without top }
\end{table}
  
The jumps of the Higgs length expectation values   $\langle {\rho^2} \rangle$ and $\langle
{\rho^4} \rangle$ at $T_c$ are connected to the RG
invariant discontinuities of the quadratic and quartic continuum Higgs
condensates  (Table~2)
\begin{eqnarray*}
\label{eq:scalar_jumps}
\Delta \langle \phi^+ \phi \rangle/g_3^2 &  = &  {1 \over 8} \beta_G \beta_{Hc}
\Delta \langle {\rho^2} \rangle,  \\
\Delta \langle (\phi^+ \phi)^2 \rangle/g_3^4  & = &  ({1 \over 8} \beta_G
\beta_{Hc})^2  \Delta \langle {\rho^4} \rangle .
\end{eqnarray*}
\begin{table}[!htb]
\begin{tabular*}{75mm}{@{\extracolsep{\fill}}|c|r|r|}
\hline
 ($\beta_G,M_H^*$) & $\Delta \langle \phi^+ \phi \rangle/g_3^2$
                   & $\Delta \langle (\phi^+ \phi)^2 \rangle/g_3^4$ \\
\hline
$(12,70)$  &  0.250(3)    &     1.28(2)  \\ 
$(16,70)$  &  0.250(4)    &     1.65(3)  \\ 
$(12,35)$  &  3.20(1)     &     25.2(1)  \\ 
\hline
\end{tabular*}
Table~2. {\small \sl The Higgs condensate discontinuities } 
\end{table}

$\Delta \langle \phi^+ \phi \rangle/g_3^2$ at $M_H^*=70$ GeV is already 
independent of finite $a$
effects. On the contrary, $\Delta \langle (\phi^+ \phi)^2 \rangle/g_3^4$ 
 shows a severe $a$
dependence which makes it   more subtle to extract an
appropriate continuum value.

 The latent heat $L_{heat} $   
is calculated according to \cite{FarakosEA} 
\begin{eqnarray*}
\label{eq:lat_heat}
 \frac{L_{heat}}{T_c^4} &=& \frac{M_H^2}{T_c^3} \   \Delta
\langle {\phi^+\phi} \rangle .
\end{eqnarray*}
With the reported numbers we
find for $L_{heat}/T_c^4$  0.0178(3) at $M_H^*=70$   and 
0.183(1) at $M_H^*=35$ GeV (without fermions). Taking into account the top
($m_t=175$ GeV) the first
number increases to 0.0574(9)
 ($T_c=107.05$ GeV, $M_H=69.42$ GeV).
 \begin{figure}[!bth]
\centering
\epsfig{file=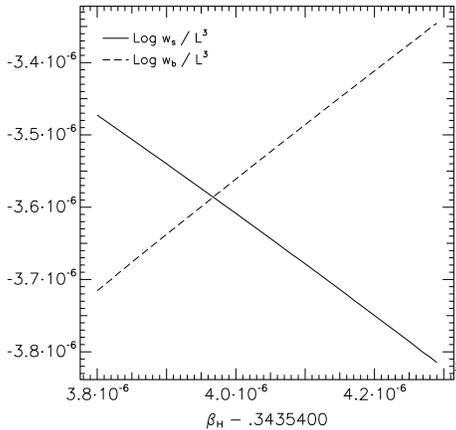,width=5.7cm,angle=90}
\vspace*{-1.1cm}
\caption{\small \sl Weights for the pure phases  
  vs. $\beta_H$ for $\beta_G=12$, $64^3$}
\label{fig:weights_12_64}
\end{figure}

 The equal weight method allows to reconstruct directly the
free energy densities of the pure phases  
near to $T_c$.   $L_{heat}$ can then be expressed
alternatively   by ($w_{s/b}$  weights of symmetric/broken phase)
$$
 \frac{L_{heat}}{T_c^4} =
- \ {g_3^2 \over 8 T_c^3 L^3 }\   M_H^2 \beta_{Hc}^2 \beta_G\
 {\partial \over \partial \beta_H}
 \log {w_s \over w_b} \Big|_{\beta_{Hc}}.
$$
 The change of the weights with $\beta_H$ very close to the critical one is
shown in Fig.~\ref{fig:weights_12_64}.
  We obtain (without top) $L_{heat}/T_c^4 = 0.0182(9)$.

The coexistence of both phases  opens the
possibility to determine the surface tension $\alpha$.  The use of the
equal weight method  allows to estimate the
contribution of the mixed phase state  
at $w_b=w_s$.

We parametrize the relation between the weights at pseudo-criticality
and $\alpha$ for lattices   $L_x^2 \times L_z$ as follows
$$
  {{w_{mix}}\over {w_s}}= {{w_{mix}} \over{w_b}}= b \; L_z^2 \; \log L_x
\exp(-2\alpha a^2 L_x^2/T_c).
$$
$L_z^2$ is an entropy factor      
(surface positions), $\log L_x$ is the result (for $d=3$) of
 capillary wave approximation (fluctuations of the
surfaces), 
 $b$   counts different
possible orientations of  surfaces (cubic ($b=3$)
and   prolongated lattices ($b=1$)).
\begin{figure}[!thb]
\centering
\epsfig{file=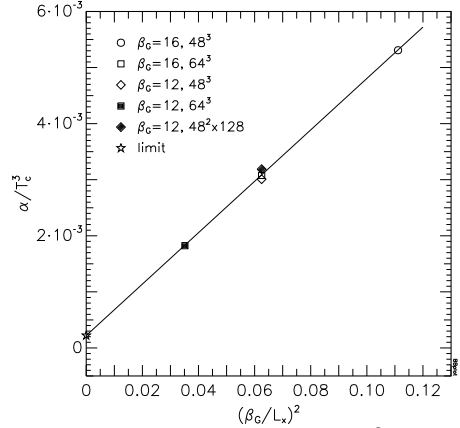,width=5.7cm,angle=90}
\vspace*{-1.1cm}
\caption{\small \sl Surface tension vs. $(\beta_G/L_x)^2$}
\label{fig:surfacetension}
\end{figure}
  We find the upper bound (Fig.~\ref{fig:surfacetension})
$ 
 \alpha / T_c^3  \approx 0.00023 
$. 
  
These lattice results are smaller by one order of magnitude than the
one--loop estimate  for $\alpha$ \cite{KripfganzEA}. 
The surface tension is sensitive to the shape
of the effective potential in the whole $\varphi$ range between the symmetric
and the broken phase. The disagreement seems to
indicate that the loop expansion to the effective potential gets out of
control at intermediate $\varphi$ values already, reflecting the infrared
problems of the symmetric phase.
\vspace*{-5mm}
\section{Broken phase and perturbation theory}
\vspace*{-2mm}
Can the broken phase   be understood perturbatively? 
It is known \cite{KripfganzEA} that   the appropriate effective
expansion parameter   is
$g_{3\,eff}^2= {g_3^2}/(2 m_W(T))$ with the $3d$  gauge boson mass 
$ m_W(T)$.  This coupling can be determined from the two--loop effective 
potential. Using Feynman gauge and $\mu_3=g_3^2$ for $M_H^*=70$ GeV this 
effective coupling is found to be $1.10$ at $T_c$.

\begin{figure}[!thb]
\centering
  \epsfig{file=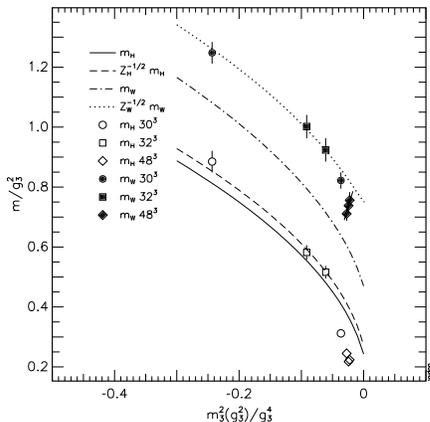,height=5.7cm,angle=90} 
\vspace*{-1.1cm}
 \caption{\label{brophase1} \small \sl $3d$ masses at $M_H^*=70$ GeV and $\beta_G=12$
 compared to   perturbation theory}
 \end{figure}

We have compared our MC data with two--loop continuum predictions 
(Feynman gauge) for  the
vector boson ($m_W(T)$) and Higgs boson ($m_H(T)$) masses (shown
in Fig.~\ref{brophase1}) and the renormalized Higgs condensate
using  the effective potential \cite{KripfganzEA}.
The $3d$ masses are calculated from correlators of extended operators with 
appropriate quantum numbers.
 
\vspace*{-3mm} 
Very good agreement is observed deeper in the broken phase, as 
should be expected.
Wave function renormalization (in one--loop) is obviously required.
Close to the phase transition we observe a 
systematic difference between lattice data and the continuum
calculation as function of $m_3^2$.
This may be an indication that higher loop terms
start to play a significant role. 
\vspace*{-3mm}
\section{Some properties of the symmetric phase}
\vspace*{-2mm}
Dynamical Higgs fields are expected to screen the static potential
which can be defined in the $3d$ theory, too, and will be described below.
Sufficiently away from the
transition massive Higgs bound states become heavy.
As a consequence, the symmetric phase
should more and more resemble pure $SU(2)$ gauge theory. 
However, in the $\beta_H$ (resp.  $m_3$) region that
we have explored the lowest
Higgs bound state is still significantly lighter than the
lightest $0^{+} \ W$--ball.
 
\begin{figure}[!thb]
\centering
\epsfig{file=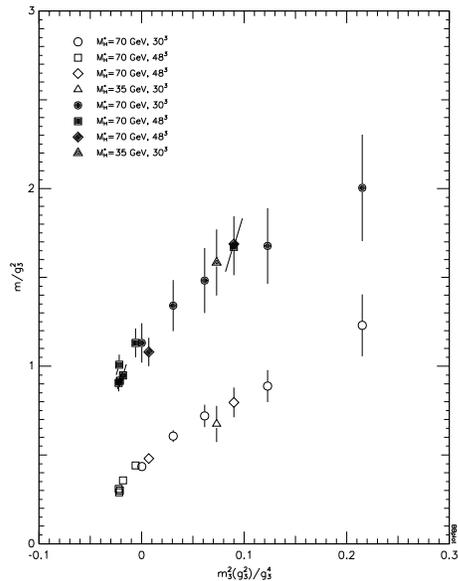,width=6.0cm,angle=0}
\vspace*{-1.1cm}
\caption{\small \sl Masses in the symmetric phase for $\beta_G=12$, diamonds
 correspond to $\beta_G=16$}
\label{fig:H_and_Wmass}
\end{figure}

Results for the lowest Higgs   ($0^{+}$) and   vector boson
bound states ($1^{-}$) are shown in Fig.~\ref{fig:H_and_Wmass} as
 function of $m_3(g_3^2)/g_3^4$. Data from different $\beta_G$ values
nicely coincide. Our results for  $m_W$ should be
considered as upper bounds, because of a possible admixture of states
with a somewhat higher mass. One important conclusion from 
Fig.~\ref{fig:H_and_Wmass} is the
scaling behavior of masses (no $\lambda_3$ dependence). 

 %
%
%
%
%

  The data for the static potential $V(R)$ are obtained from exponential
fits to the Wilson loops.
The potential (in $g_3^2$ units) is described assuming massless or
 massive perturbative $W$--exchange.
 
\begin{figure}[!htb]
\centering
\epsfig{file=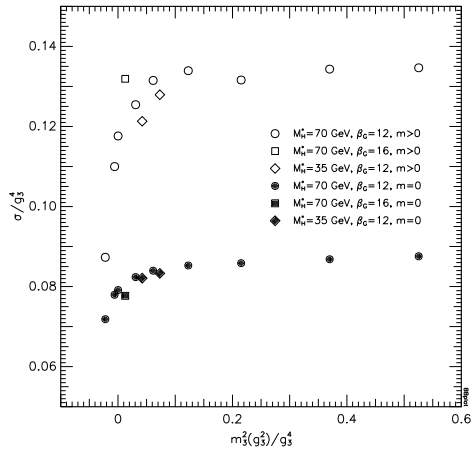,width=6.0cm,angle=90}
\vspace*{-1cm}
\caption{\small \sl String tension $\sigma/g_3^4$  vs. $m_3^2(g_3^2)/g_3^4$}
\label{fig:st}
\end{figure}

In Fig.~\ref{fig:st} we present the string tension obtained from the
ans\"atze for the potential   at different $m_3^2$ values.
 There is no significant $\lambda_3$ ($M_H^*$)
dependence of $\sigma/g_3^4$ on $\lambda_3$ ($M_H^*$). 
   At larger $m_3^2$ (temperatures)  the result of the string  
tension based on the fit to
massive perturbation theory   is  
close to the value   of pure $3d$ $SU(2)$.  
For  $R \le 18a$   the expected screening
behavior of the  potential has not yet been observed. Still
larger distances are difficult to study. 
\vspace*{-5mm}
\section{Summary}
\vspace*{-2mm}
Our results   provide evidence for the first
order nature of the  thermal phase transition in the $SU(2)$--Higgs
system with Higgs masses up to $70$ GeV. 

We checked the reliability of perturbative calculations
of the effective potential (masses and   renormalized
Higgs condensate in the broken phase). The effect
of wave function renormalization cannot be ignored.  Physics depending
  only on the potential in the vicinity of the broken minimum,
can be systematically improved by higher order perturbation theory.

The surface tension which is sensitive to the barrier shape of the
effective potential  
is systematically overestimated in the perturbative calculation.
    It is also very hard to
measure for weak transitions. 

  The reduced model does not only make very
precise predictions, but   dimensional reduction 
seems to be reliable in the interesting range of Higgs
masses.  
In order to learn about the necessity to include higher dimensional 
operators into the effective  action one should explore the
case of smaller or very much larger Higgs masses.
 The $3d$
lattice approach promises to be applicable as an effective formulation
of nonstandard extensions of the Standard Model. 

We have put a lot of emphasis   to the properties of the
symmetric phase. Within the $3d$ approach, information on the spectrum
 can only be accessed through   $3d$
correlation lengths. 
The interplay between the confining properties
of the $3d$ effective theory at high $T$ and the
space--time structure of physical excitations needs further investigations, 
as well as the nature of this confinement itself for the $3d$ pure gauge 
theory and in the presence of scalar matter fields.

\end{document}